\RequirePackage{lineno}
 \documentclass[aps,prd,twocolumn,preprintnumbers,superscriptaddress,nofootinbib]{revtex4}
\usepackage{amsmath}
\usepackage{colortbl}
\usepackage{xcolor}
\usepackage{graphicx}  
\usepackage{dcolumn}   
\usepackage{bm}        
\usepackage{amssymb}   
\usepackage{multirow}
\usepackage{lineno}
\usepackage[hyperfootnotes=false]{hyperref}
\usepackage{array, graphicx, subfigure}
\usepackage{graphics}
\begin{document}

\newcommand{\carbon}{\rm ^{12}C}
\newcommand{\oxygen}{\rm ^{16}O}
\newcommand{\deuteron}{\rm ^{2}H}
\newcommand{\hydrogen}{\rm ^{1}H}
\newcommand{\Hefour}{\rm ^{4}He}
\newcommand{\lead}{\rm^{208}Pb}
\newcommand{\Hethree}{\rm ^{3}He}
\newcommand{\neon}{\rm ^{20}Ne}
\newcommand{\aluminum}{\rm^ {27}Al}
\newcommand{\argon}{\rm ^{40}Ar}
\newcommand{\iron}{\rm ^{56}Fe}
\newcommand{\genie}{$\textsc{genie}$}
\newcommand{\gep}{$G_{Ep}$ } 
\newcommand{\gen}{$G_{En}$ } 
\newcommand{\qv}{$\bf |\vec q|$}
\newcommand{\rlqe}{$R_L^{QE}(\bf q, \nu)$ }
\newcommand{\rtqe}{$R_T^{QE}(\bf q, \nu)$ }
\newcommand{\rltot}{$R_L(\bf q, \nu)$ }
\newcommand{\rttot}{$R_T(\bf q, \nu)$ }
\newcommand{\Rochester}{Department of Physics and Astronomy, University of Rochester, Rochester, NY  14627, USA}
\newcommand{\JLAB}{Thomas Jefferson National Accelerator Facility, Newport News, VA 23606, USA}
%
\title{Extraction of the Coulomb Sum Rule, Transverse Enhancement, and Longitudinal Quenching from an Analysis of all Available e-$^{12}$C and e-$^{16}$O Cross Section Data.
} 
\affiliation{\Rochester}
\affiliation{\JLAB}
  \author{A.~Bodek}
\affiliation{\Rochester}
    \author{M.~E.~Christy}
\affiliation{\JLAB}
 \
\date{\today}
\begin{abstract}
 We report on a phenomenological analysis of  all available electron scattering data  on ${\rm ^{12}C}$ (about 6600 differential cross section measurements) and  on ${\rm ^{16}O}$ (about 250 measurements) within the framework of the quasielastic (QE) superscaling  model (including Pauli blocking).  All QE and inelastic cross section measurements  are included down to the lowest momentum transfer  $\bf q$ (including photo-production data). We find that there is enhancement of the transverse QE response function ($R_T^{QE}$) and quenching of the QE longitudinal response function ($R_L^{QE}$) at low  $\bf q$ (in addition to Pauli blocking). 
 We extract parameterizations of a  $multiplicative$ low $\bf q$ "Longitudinal Quenching Factor" and an $additive$ "Transverse Enhancement" contribution.   
 Additionally, we find that the excitation of  nuclear states contribute significantly (up to 30\%) to the Coulomb Sum Rule $SL({\bf q})$. We extract the most accurate determination of $SL({\bf q})$ to date and find it to be in disagreement with   random phase approximation (RPA)
 based calculations but in  reasonable agreement with recent theoretical calculations such as "First Principle Green's Function Monte Carlo".
\end{abstract}
\maketitle 
We report on a fit to all available electron scattering data on $\carbon$ (about 6600 differential cross section measurements) and $\oxygen$ (about 250 measurements) within the framework of the  quasielastic (QE) superscaling  model (including Pauli blocking).  The cross sections measurements include the available data on QE (down to the lowest momentum transfer $\bf q$ ($\equiv |\vec q|$), inelastic production, and photoproduction.
The fit includes inelastic structure functions and empirical parameters to model both an enhancement of the transverse (T) QE response function $R_T^{QE}$ and  quenching of the  longitudinal (L) QE  response  function  $R_L^{QE}$ at low  $\bf q$. As the fit provides an accurate description of the data, it can be used as a proxy to validate modeling of cross sections in  Monte Carlo event generators for electron and neutrino ($\nu_{e,\mu}$) scattering.  Careful consideration of nuclear excitations is critical for an accurate extraction of the normalized Coulomb Sum Rule\cite{CSR} $SL({\bf q})$ at low $\bf q$ as these states can contribute up to 30\%.   After accounting for the dominant excitations, we extract the most accurate determination of $SL({\bf q})$) as function of $\bf  q$ for $\carbon$ and $\oxygen$ based on the global fit and compare to theoretical models. In addition, the "Transverse Enhancement" (TE) of $R_T^{QE}$ and the "Quenching Factor"  of $R_L^{QE}$ are also of great interest to $\nu_{e,\mu}$ scattering experiments\cite{miniboone,minerva,microboone1,microboone2}. 
 
 
The electron scattering differential cross section can be written\cite{GEp} in terms of  $R_L({\bf q},\nu)$ and $R_T({\bf q},\nu)$ as: 
%
\begin{eqnarray}
\frac{d^2\sigma}{d \nu d\Omega}&=& \sigma_M [ A R_L ({\bf q},\nu) + B R_T ({\bf q}, \nu)],\\
\sigma_M &=& \alpha^2 \cos^2(\theta/2)/[4 E_0^2 \sin^4(\theta/2)].\nonumber
\end{eqnarray}
%
%
Here,  $E_0$ is the incident electron energy,  $E^{\prime}$ and $\theta$ are energy and angle of the final state electron, 
$\nu=E_0-E^{\prime}$, 
$Q^2$ is the square of the 4-momentum transfer (defined to be positive),  $\bf {q}^2$=$Q^2+\nu^2$,  $A=(Q^2/\bf {q}^2)^2$  and $B = \tan^2(\theta/2) +Q^2/2\bf{q}^2$. In the analysis we also use the invariant hadronic mass $W^2= M_p^2+2M_p\nu-Q^2$. 

The inelastic Coulomb Sum Rule is the integral of  $R_L({\bf q},\nu)d\nu$, {\it excluding the elastic peak and  pion production processes}. It has contributions from QE scattering and from electro-excitations of nuclear states: 
\begin{eqnarray}
\label{CSReq}
&&{\rm CSR}({\bf q})= \int R_L({\bf q},\nu)d\nu\\
&=& \int R_L^{QE}({\bf q},\nu) d\nu\nonumber
+ G^{\prime 2}_E(Q^2) \times Z^2 \sum_{all}^{L} F^2_i({\bf q})\\\nonumber
&=& G^{\prime 2}_E(Q^2) \times \big[ Z \int V^{QE}_L({\bf q},\nu) d\nu + Z^2 \sum_{all}^{L} F^2_i({\bf q})\big]\nonumber.
\end{eqnarray}
%
We define $V^{QE}({\bf q},\nu)$ as the reduced longitudinal QE response, which integrates to unity in the absence of any suppression (e.g. Pauli blocking).
The charge form factors for the electro-excitation of nuclear states $F^2_{iC}({\bf q})$ is $G^{2}_{Ep} (Q^2)\times F^2_i({\bf q})$. 
In order to account for the  small contribution of the neutron and relativistic effects $G^{\prime 2}_E(Q^2)$ is given by\cite{GEp}:
 \begin{equation}
 G^{\prime 2}_E(Q^2)= [G^2_{Ep}(Q^2)+\frac{N}{Z}G^2_{En}(Q^2)] \frac{1+\tau}{1+2\tau},
 \label{GEprime}
  \end{equation}
where, \gep and \gen are the electric form factors~\cite{BBBA} of the proton and neutron respectively and  $\tau=Q^2/4M_p^2$. 
%
By dividing Eq.~\ref{CSReq} by $ZG^{\prime 2}_E {\bf q})$ we obtain the normalized inelastic Coulomb Sum Rule $SL({\bf q})$ :
\begin{eqnarray}
\label{SLINE}
SL({\bf q})= \int V^{QE}_L({\bf q},\nu)d\nu + Z \sum_{all}^{L} F^2_i({\bf q}).
\end{eqnarray}

At high $\bf q$ it is expected that $S_L \rightarrow 1$ because both nuclear excitation form factors and Pauli suppression are small. At small $\bf q$ it is expected that $S_L \rightarrow 0$ because the all form factors for inelastic processes (QE and nuclear excitations) must be zero at $\bf q$=0.  

We begin by parameterizing the measurements of the  L and T form factors for the electro-excitation of all  nuclear states in $\carbon$ with excitation energies ($E_x$) less than 16.0 MeV (the approximate proton removal energy  from $\carbon$). For these states the measurements are straightforward since the QE cross section is zero for $E_x<$ 16 MeV.  

\begin{figure}[ht]
\includegraphics[width=3.4in, height=1.5 in]{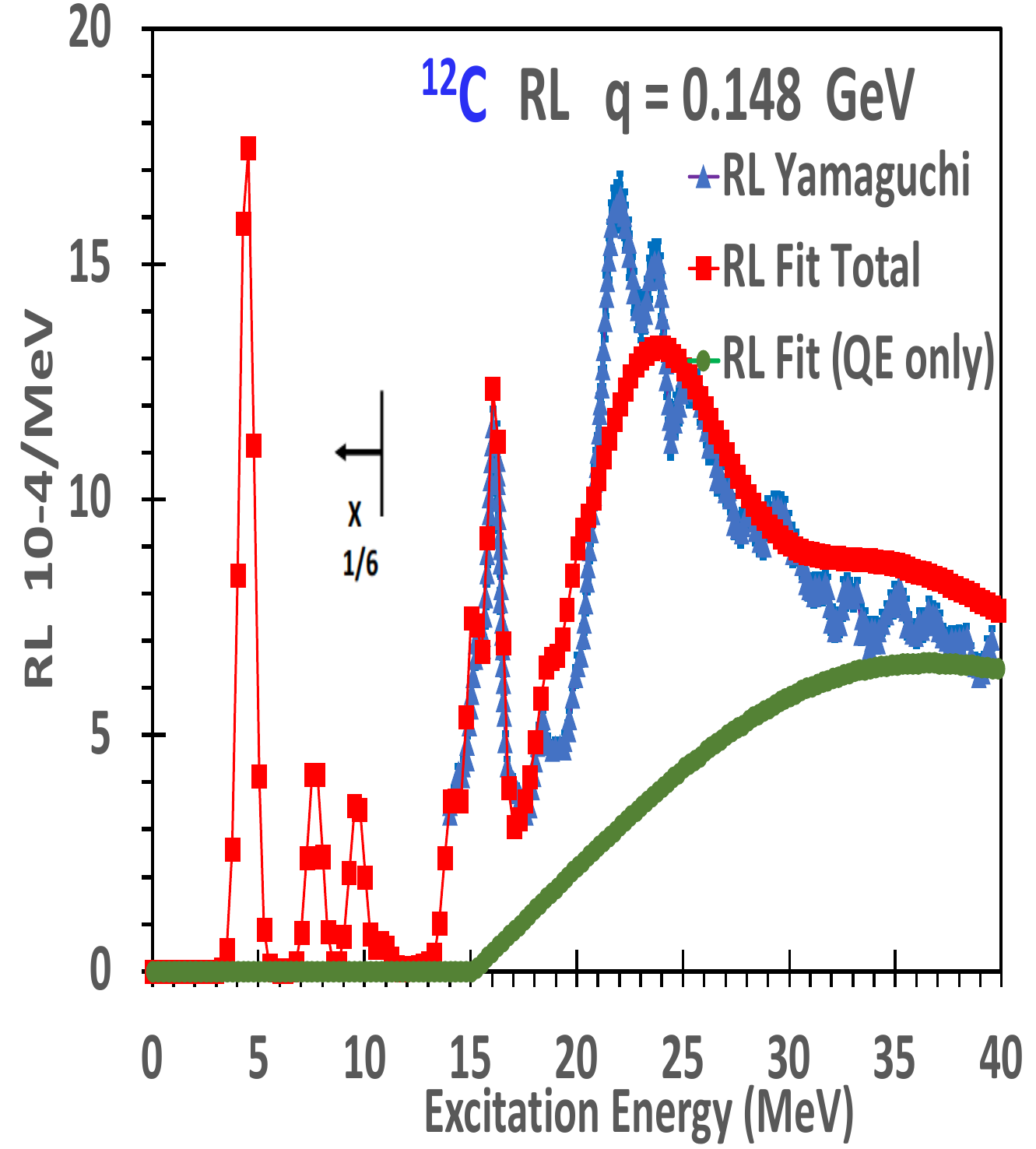}
\includegraphics[width=3.4in, height=1.5 in]{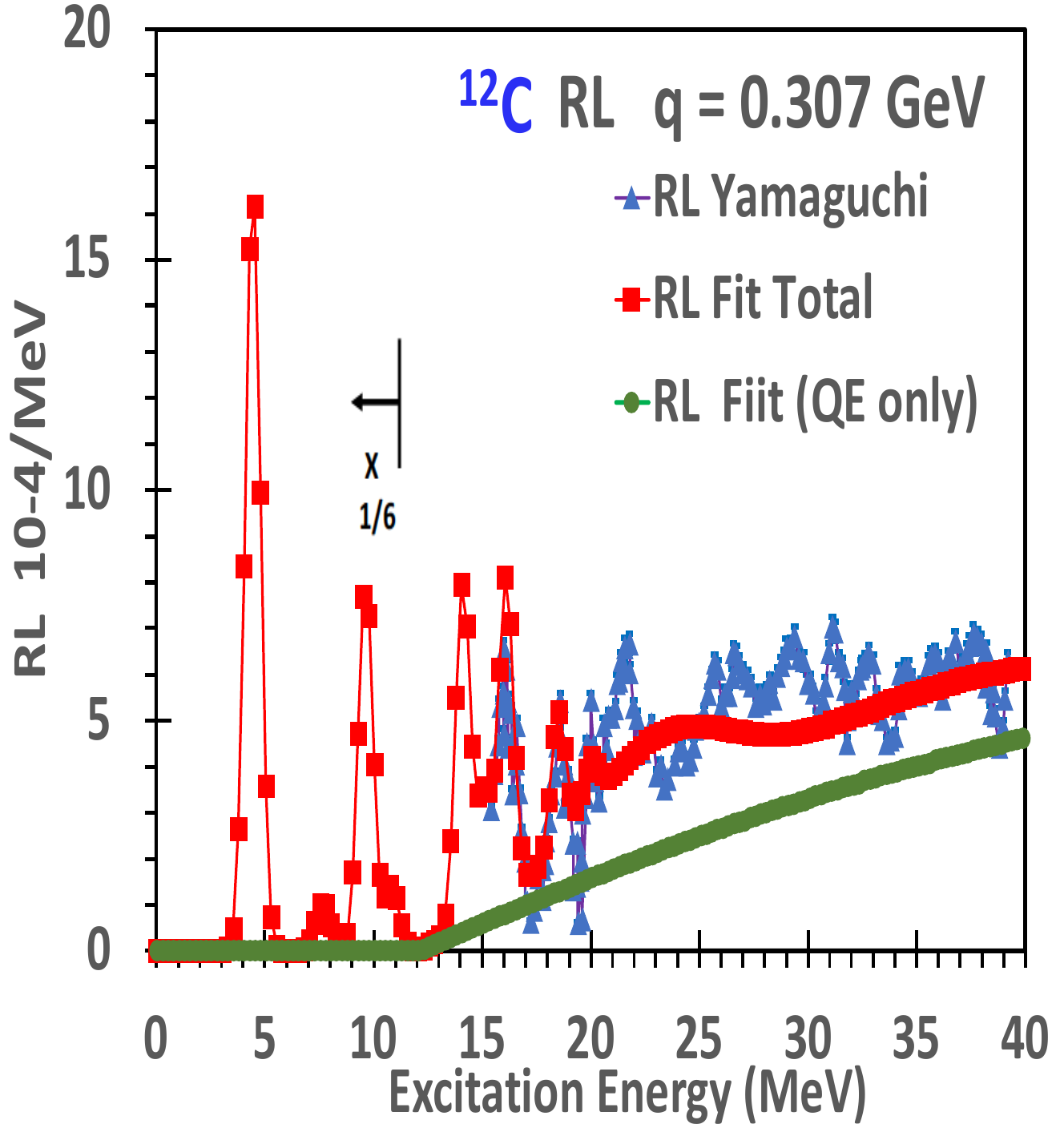}
\includegraphics[width=3.4in, height=1.8 in] {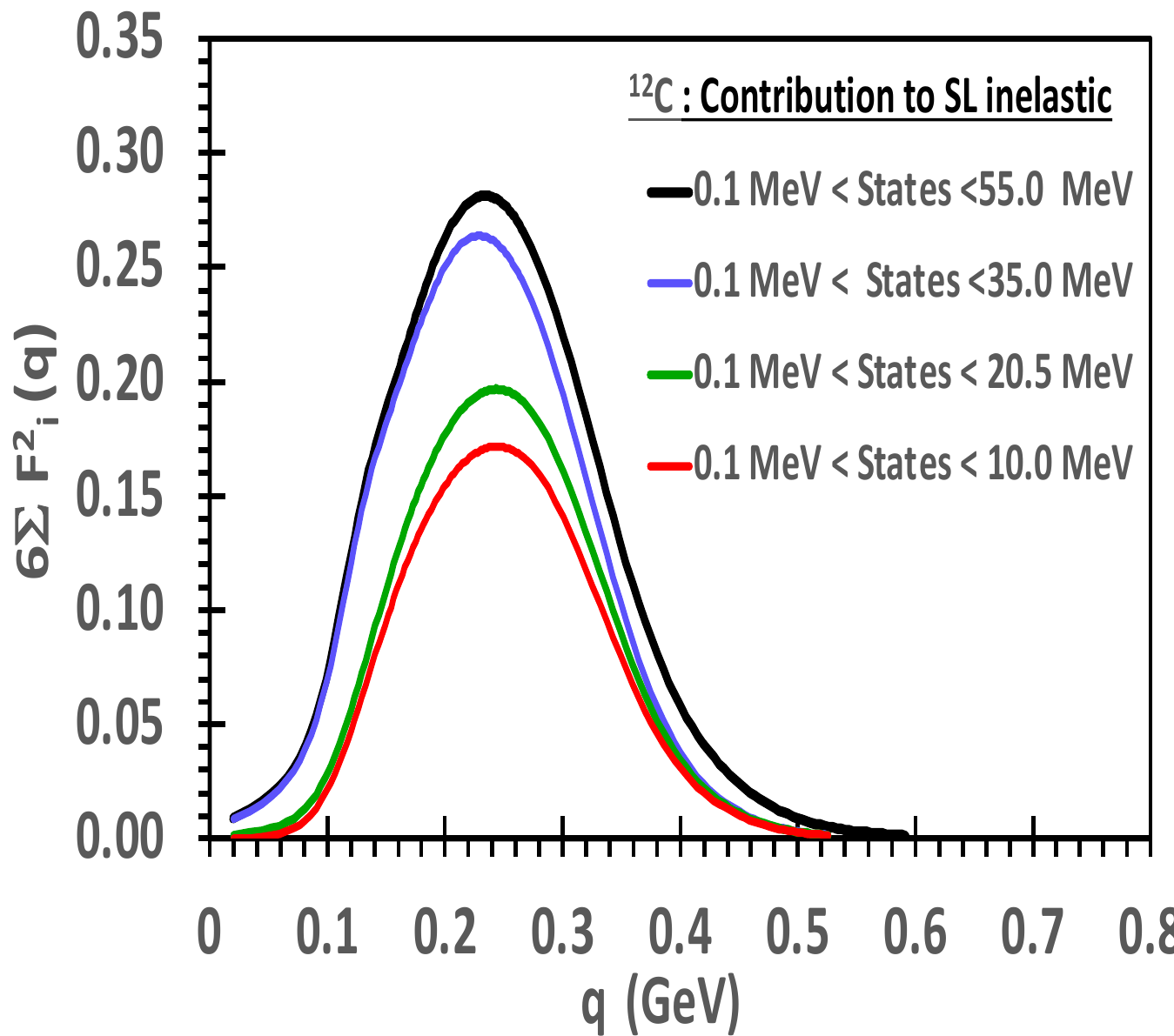}
\includegraphics[width=3.4in, height=1.8in] {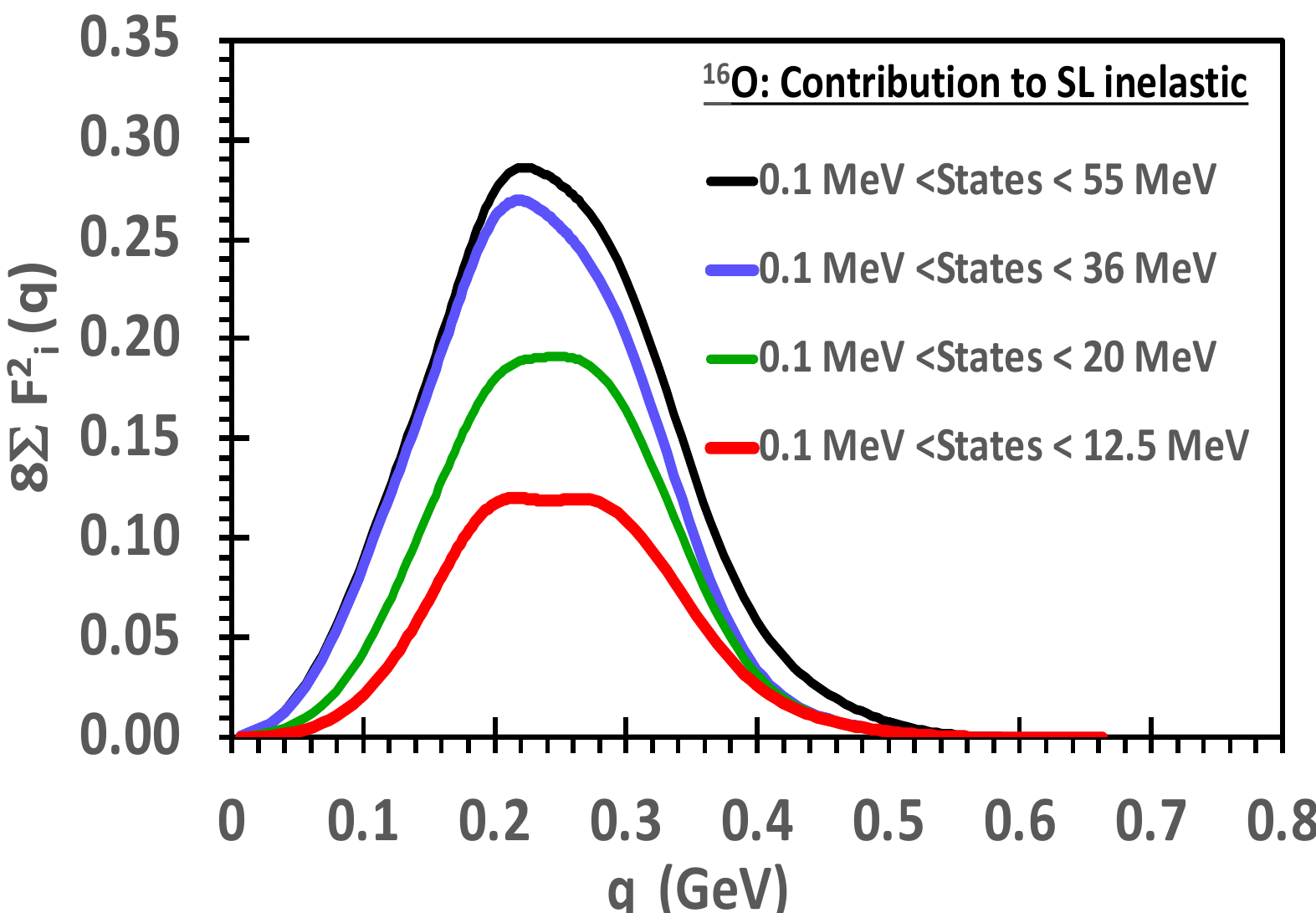}
\vspace{-10pt}
\caption{Top two panels: Comparison of  $R_L({\bf q}, \nu)$  extracted from our $\carbon$ fit (red) to a sample of  experimental data (blue)\cite{Yamaguchi}. 
For $E_x$ less than 12 MeV  the values are multiplied by 1/6. Bottom two panels: The contributions of longitudinal nuclear excitations (between 2 and 55 MeV) to the  Coulomb sum rule ($Z \sum_{all}^{L} F^2_i({\bf q})$) in equation  \ref{SLINE} for $\carbon$  and $\oxygen$.}
\vspace{-15pt}
\label{Fig1}
\end{figure}

For $E_x>$~16 MeV the extractions of form factors require corrections for the QE contribution. We perform a reanalysis of all published cross sections with in $E_x<$~55 MeV and use our fitted QE model (described below) to extract  L and T form factors.  For $E_x>$~20 MeV (region of the Giant Dipole resonances) we group the strength from multiple excitations into a few states with a large width $E_x$ and extract effective form factors accounting for all states in this region.  The top two panels of  Fig. \ref{Fig1} show comparisons of our fit (red) to $R_L$ measurements by Yamaguchi (blue) 1996\cite{Yamaguchi} with $E_x>$~14 MeV. An estimated resolution smearing of 600 keV has been applied to the excitations in the fit to match the data.
While individual states are well reproduced at low excitation energy, Above 20~MeV the effect of 
grouping several excitations together into broad effective states in the fit can be seen.  While the fit does not capture the structure from individual states, the total strength is seen to be well reproduced.  A similar analysis has been done for $\oxygen$. The fits to the form factors for $\carbon$) and 
$\oxygen$ are included in a longer paper (in preparation).

The contribution of nuclear excitation to $SL({\bf q})$ (factor  $Z \sum_{all}^{L} F^2_i({\bf q})$ in eq.~\ref{SLINE}) is calculated using the parametrizations of the form factors.  The bottom two panels of Fig.~\ref{Fig1} show the contributions of nuclear excitations to
$SL({\bf q})$ for $\carbon$ and  $\oxygen$.  The contribution of all excitations is largest ($\approx$ 0.29) at $\bf q$=0.22 GeV.  
Although the contributions of different $E_x$ regions to  $SL({\bf q})$ is different for $\carbon$ and $\oxygen$, the total contribution turns out to be similar for the two nuclei. 
The  total contribution of excitations to $S_L({\bf q})$ in $\carbon$ can be  parameterized as: 
\begin{eqnarray}
\label{all states}
&Z&\sum_{all}^{L} F^2_i({\bf q}) =N_1 e^{-(x-C_1)^2/D_1^2}\\
&+& N_2 e^{-(x-C_2)^2/D_2^2} 
+ N_3 e^{-(x-C_3)^2/D_3^2}\nonumber
\end{eqnarray}
where  x= ${\bf q}/K_F$ ($ K_F$= 0.228 GeV), $N_1$= 0.260, $C_1$=1.11,  $D_1$=0.50, $N_2$= 0.075, $C_2$=0.730,  $D_2$=0.30, and 
$N_3$= 0.01, $C_3$=2.0,  $D_3$=0.30.
  The uncertainty in the total contribution of  excited states was estimated to  15\%  plus a systematic error to account for the choice of parametrization at very low $\bf q$ ($\pm$0.01) added in quadrature.
%
%
%
%

\begin{figure*}
 %
\includegraphics[width=7.0in, height=3.6in]{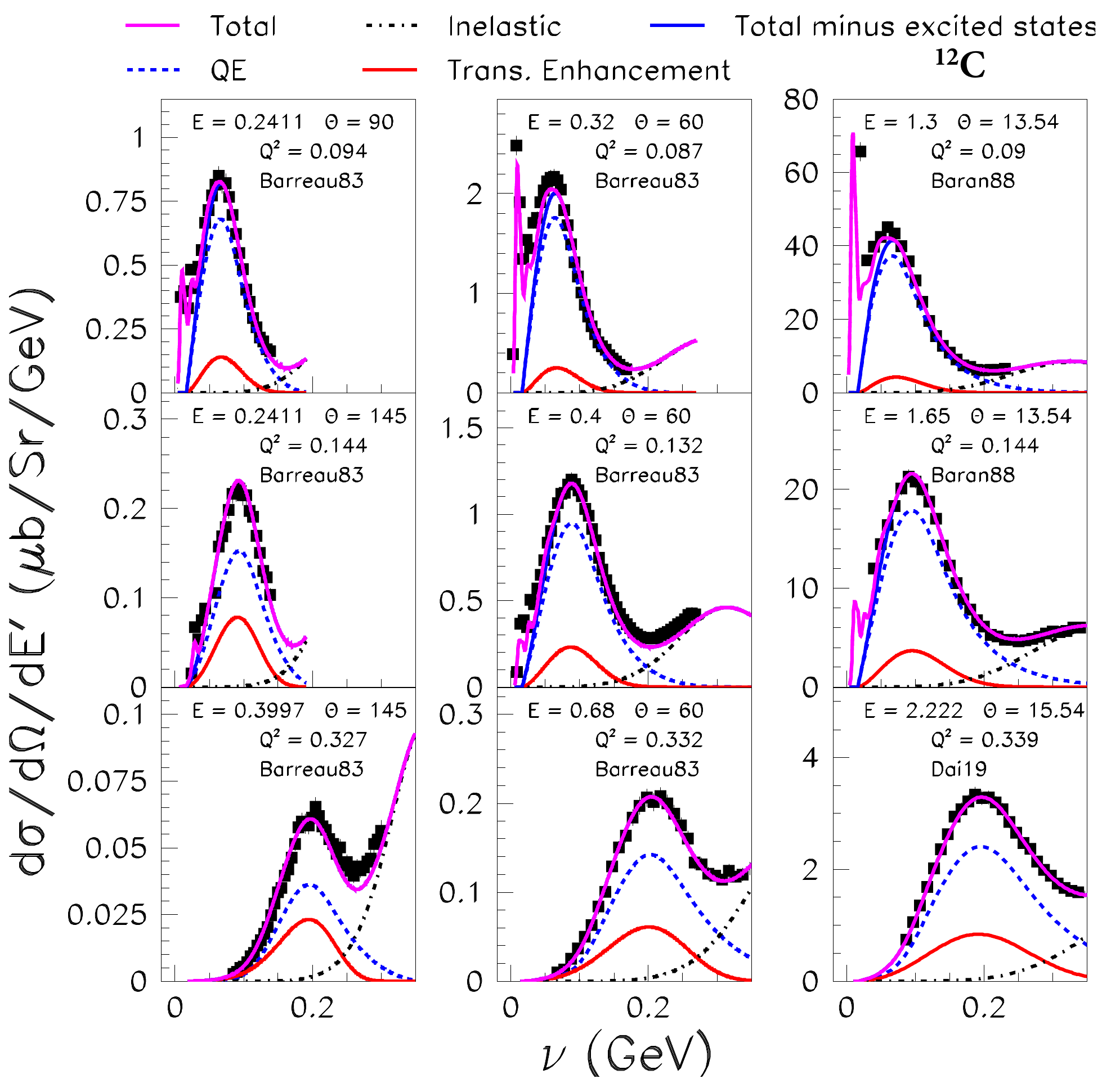}
\vspace{-10pt}
 \caption{Comparison of  the fit to electron scattering $\frac{d^2\sigma}{ d\Omega d \nu}$ measurements\cite{archive,Barreau,Baran} at $\bf q$ values close to  0.30, 0.38 and 0.57 GeV (and different scattering angles).
  Shown are total $\frac{d^2\sigma}{ d\Omega d \nu}$ (solid-purple line), total minus the contribution of the nuclear excitations (solid-blue), the QE cross section without TE (dashed-blue),  the TE contribution (solid-red) and  inelastic pion production (dot-dashed black line). 
 Additional comparisons are included in supplemental materials\cite{Supplemental}.
 }
 \vspace{-10pt}
\label{Fig2}
\end{figure*}
The universal fit to the $\carbon$ data  is an update   of  the 2012 fit by Bosted and Mamyan \cite{Bosted}.
The  QE contribution  is modeled by the superscaling approach\cite{Donnelly, Amaro1, Amaro2, Megias}  with Pauli blocking calculated using the Rosenfelder\cite{Megias, Rosenfelder,Megias2} method. The superscaling function extracted from the fit is similar to the superscaling functions of  Amaro 2005\cite{Amaro1} and Amaro 2020\cite{Amaro2} and yields similar Pauli suppression.

In modeling the QE response we use the same scaling function for both $R_L^{QE}({\bf q},\nu)$ and $R_T^{QE}({\bf q},\nu)$ 
and fit for empirical corrections to the
response functions. 
 For  $R_T^{QE}$ we extract an {\it additive} "Transverse Enhancement/MEC" TE(${\bf q},\nu$)  contribution (which includes both single nucleon and two nucleon final states).
 As shown in  ref.~\cite{transverse}.   TE(${\bf q},\nu$) increases $R_T^{QE}$ with the largest  fractional contribution around $Q^2$=0.3 GeV$^2$.
  For  $R_L^{QE}$  we extract a {\it multiplicative} $\bf q$ dependent "Longitudinal Quenching Factor", $F_{quench}({\bf q})$, which decreases $R_L^{QE}$ at low  $\bf q$.
 
 Since $\frac{d^2\sigma}{ d\Omega d \nu}$ measurements span a large range of  $\theta$  and $\bf q$, parametrizations of  both  TE(${\bf q},\nu$) and $F_{quench}^L({\bf q})$ can be extracted. 
 The analysis includes all data for a  large range of nuclei.  However, in this paper we only include data on  $\carbon$ and $\oxygen$.   Briefly, the updated fit includes:

\begin{figure*}[ht]
 %
\includegraphics[width=3.4in, height=2.in] {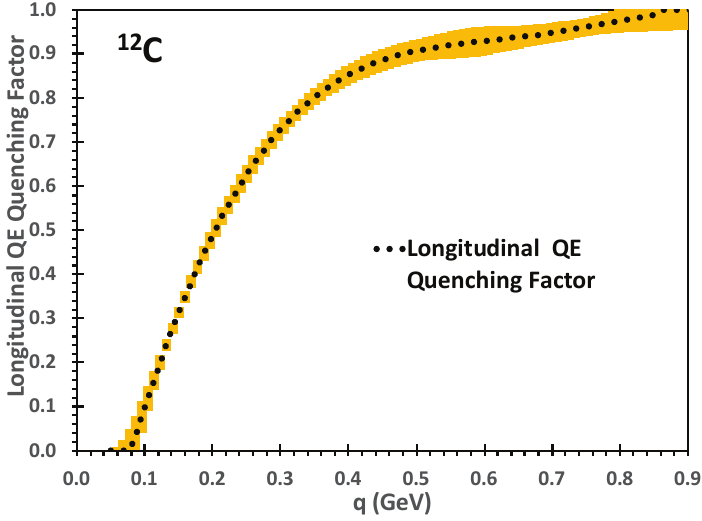}
\includegraphics[width=3.4in, height=2.in] {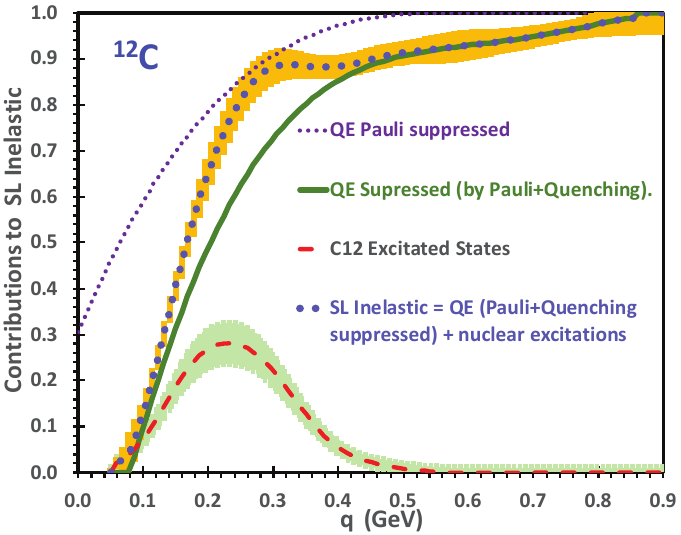}
\includegraphics[width=3.4in, height=2.in] {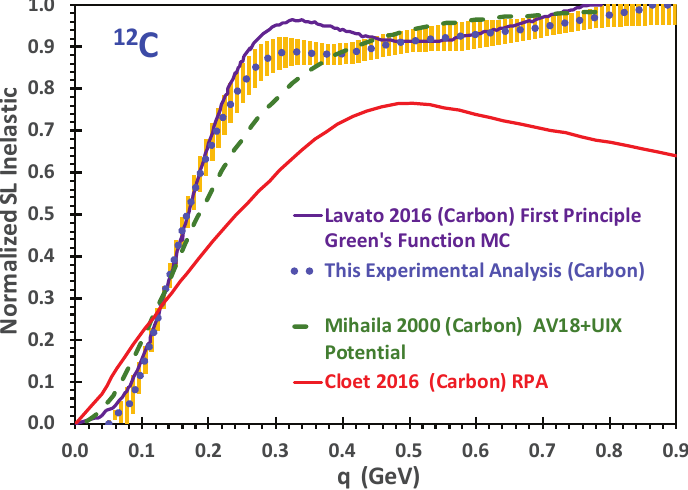}
\includegraphics[width=3.4in, height=2.in] {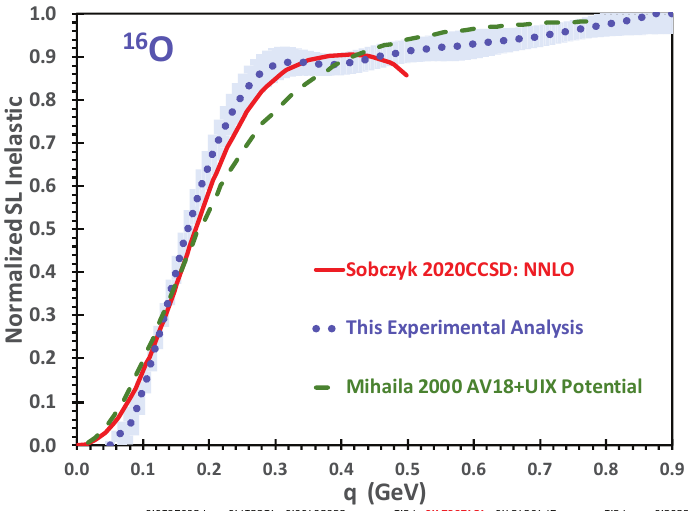}\
\vspace{-10pt}
\caption {Top left panel: QE "Longitudinal Quenching Factor" 
(dotted-black line with yellow error band). 
Top right panel: The various contributions to 
$SL({\bf q})$ for $\carbon$ (dotted blue with yellow error band) including QE  with Pauli suppression only (dotted-purple), QE suppressed by both "Pauli" and   "Longitudinal Quenching" (solid-green), and the contribution of  nuclear excitations (red-dashed with green error band).
Bottom left panel: $SL({\bf q})$ for $\carbon$ (dotted-blue with yellow error band) compared to theoretical calculations including  Lovato 2016 \cite{Lovato2016} (solid-purple), (Mihaila 2000\cite{microscopic} (dashed-green), and  RPA Cloet 2016\cite{Cloet}  (solid-red). Bottom right panel: $SL({\bf q})$ for $\oxygen$  (dotted-black with green error band) compared to theoretical  calculations of  Sobczyk 2020\cite{Coupled}  (red-dashed) and Mihaila 2000 (dotted-dashed).}
\label{Fig3}
\vspace{-10pt}
\end{figure*}
\begin{enumerate} 
\item
   All electron scattering data on $\hydrogen$, $\deuteron$, $\carbon$ and $\oxygen$ in addition to the data in the QE\cite{archive} and resonance\cite{resonance} data archives.
\item 
 Coulomb corrections\cite{Coulomb} using the Effective Momentum Approximation (EMA) in modeling scattering from nuclear targets.
\item
Updated nuclear elastic+excitations  form factors.
\item  
 Superscaling  $FN(\psi^\prime)$ parameters are re-extracted including the Fermi broadening parameter $K_F$.
\item 
 Parameterizations of the free nucleon form factors\cite{FreeN} are re-derived from all  $\hydrogen$  and $\deuteron$ data.
\item 
  Rosenfelder Pauli suppression\cite{Megias,Rosenfelder,Megias2} which  reduces and  changes the QE distribution at low $\bf{q}$ and $\nu$. 
\item 
Updates of fits\cite{FreeN} to inelastic electron scattering data (in the nucleon resonance region and inelastic continuum) for $\hydrogen$ and $\deuteron$.
\item  
 A $\bf q$ dependent  $E_{\mathrm{shift}}^{QE}(\bf q)$ parameter  for the QE process to account for the optical potential\cite{removal} of final state nucleons.   
\item 
  Photo-production data in the nucleon resonance region and inelastic  continuum\cite{ChristyD}.
\item  
 Gaussian Fermi smeared nucleon resonance and inelastic continuum\cite{ChristyD}. The $K_F$ parameters for pion production and QE can be different.
\item 
 Parametrizations of the medium modifications of both the L and T structure functions responsible for the EMC effect (nuclear dependence of inelastic structure functions).  These are applied to the free nucleon cross sections prior to application of the Fermi smearing.    
\item
Parametizations of TE(${\bf q},\nu$)  and   $F_{quench}^L({\bf q})$ as described below.
\item 
QE data at {\it all values} of $Q^2$ down to  $Q^2$=0.01 GeV$^2$ ($\bf q$=0.1 GeV) (which were not included in the Bosted-Mamyan fit).
 \end{enumerate}
 The average (over $\nu$) Pauli suppression factor for $x<2.5$ ( $x ={\bf q}/K_F$, $K_F$=0.228 GeV)
 is described by:
\vspace{-5pt}
  \begin{equation}
\label{Pauli_average}
\langle F_{Pauli}^{This-analysis}({\bf q})\rangle =  \sum_{j=0}^{j=3}  k_j (x)^j.
\end{equation}
\noindent  Using the Rosenfeld method with  supercaling function used in this analysis, we find $k_o$=0.3054, $k_1$=0.7647, $k_2$=${\bf -}$0.2768 and  $k_3$=0.0328.  
The Pauli suppression factor  for $x>2.5$ is 1.0.

Comparisons of the fit to  electron scattering $\frac{d^2\sigma}{ d\Omega d \nu}$ measurements\cite{archive,Barreau,Baran}  at  different values of $\theta$ for $\bf q$ values close to  0.30, 0.38 and 0.57 GeV (corresponding to extractions of $R_L$ and $R_T$ by  Jourdan\cite{Jourdan}) are shown in Fig. \ref{Fig2}.
Shown are the total $\frac{d^2\sigma}{ d\Omega d \nu}$ cross section (solid-purple line), the total minus the contribution of nuclear excitations (solid-blue), the QE cross section without TE (dashed-blue), the  TE contribution (solid-red), and inelastic pion production (dot-dashed black).  An estimated resolution smearing of 3.5 MeV has been applied to the excitations to better match the data.

The fit is in good agreement with all electron scattering data for both small and large $\theta$.

  The extracted QE "Longitudinal Quenching Factor"  
  $F^{L}_{quench}({\bf q})$ is unity for x$>$3.75, and is zero for x$<$0.35. 
    For $0.35< x< 4.0$ it is parameterized by:
\vspace{-5pt} 
\begin{eqnarray}
\label{F_extra}
&&F^{L}_{quench}({\bf q}) =\frac{(x-0.2)^2}{(x-0.18)^2}\big[ 1.0+A_1(3.75-x)^{1.5}\nonumber\\
          &&+A_2(3.75-x)^{2.5}+A_3(3.75-x)^{3.5}\big]
\vspace{-10pt}
\end{eqnarray}
%
with $A_1$=${\bf -}$0.13152 $A_2$=0.11693, and $A_3$=$\bf-$0.03675.
The top-left panel of  Fig.  \ref{Fig3} shows the extracted $F^{L}_{quench}({\bf q})$. 
  (black-dotted line).
  The yellow band includes the statistical, parameterization and a normalization error of 2\% (all added in quadrature).

If another formalism is used to model QE scattering (e.g. RFG or spectral functions) then the  quenching factor for the  model $F^{L-model}_{quench}({\bf q})$
is given by:
\vspace{-4pt}
\begin{eqnarray}
\label{model_extra}
F^{L-model}_{quench}({\bf q})=  \frac{\langle F_{Pauli}^{This-analysis}({\bf q})\rangle}{\langle F_{Pauli}^{model}({\bf q})\rangle} F^{L}_{quench}({\bf q})
\end {eqnarray}
  The top right panel of Fig.~\ref{Fig3} shows the various contributions to the measured  $SL({\bf q})$  for $\carbon$  (dotted blue line with yellow error band). Shown are the QE contribution with only Pauli suppression (dotted-purple), QE suppressed by both "Pauli Suppression" and $F^L_{quench}({\bf q})$  labeled as QE total suppression (solid-green),  and the contribution of nuclear excitations (red-dashed line). The green error band is 15\% plus 0.01 added in quadrature.
   \begin{figure*}[ht]
\includegraphics[width=3.4in, height=3. in]{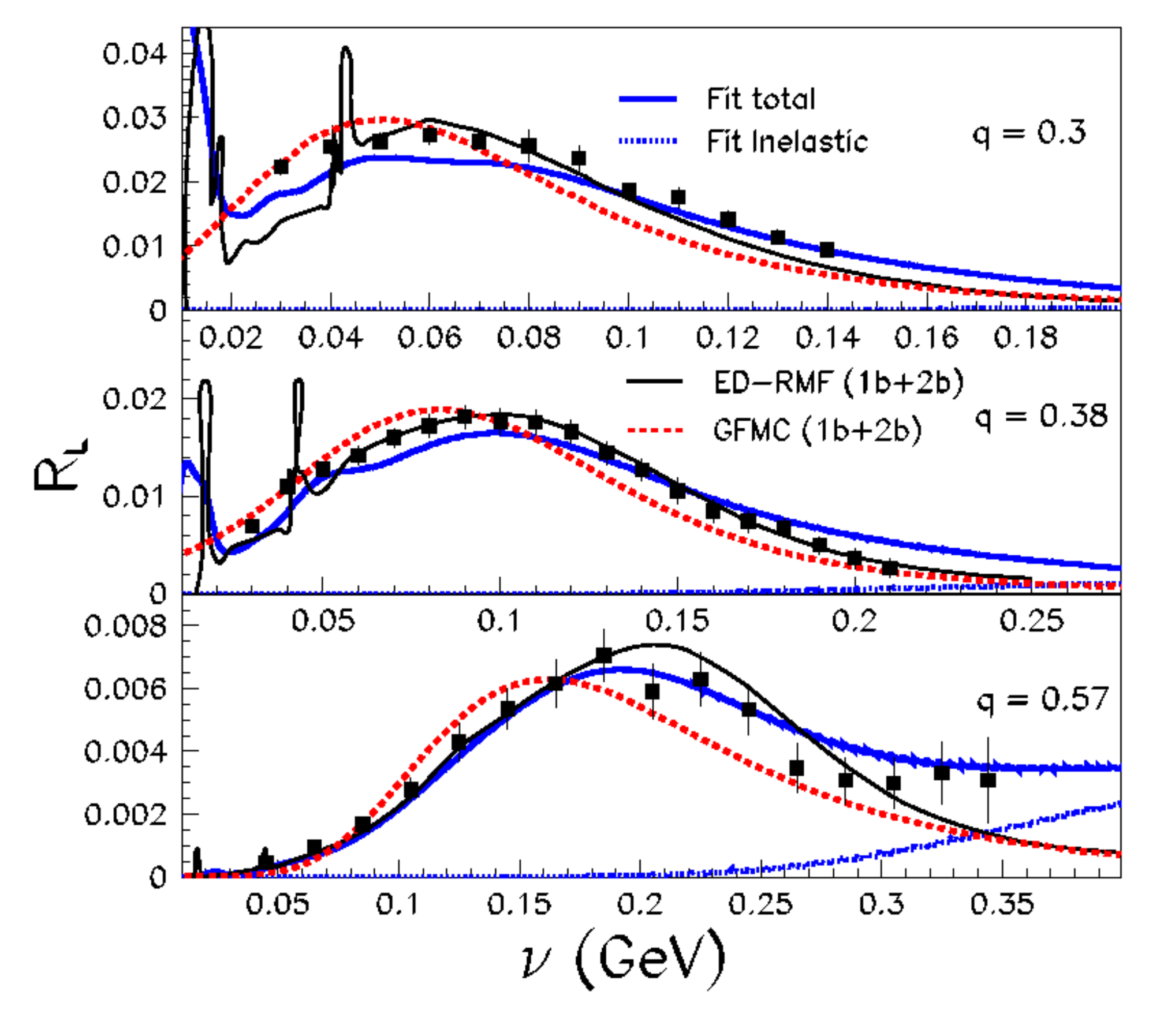}
\includegraphics[width=3.4in, height=3. in]{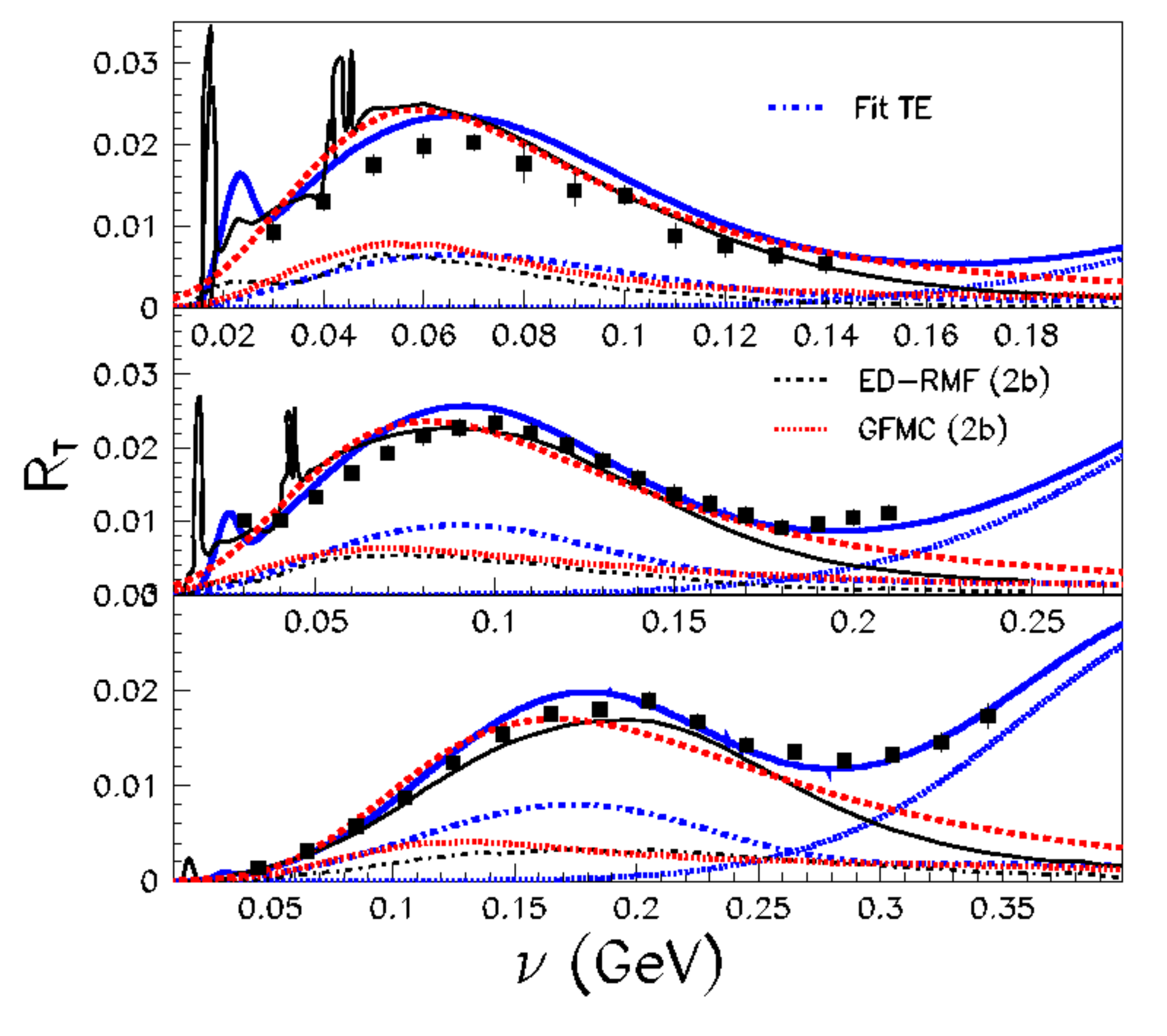}
\caption{
 Comparisons between our extraction  of \rltot and \rttot and the extraction (for only three values of $\bf q$) by Jourdan\cite{Jourdan} (the Jourdan analysis includes data from only two experiments). 
     Also shown are 
  comparisons to 1b+2b GFMC\cite{Lovato2016} and ED-RMF\cite{ED-RMF} theoretical predictions. In  these two models the curves labeled 2b are the only contribution of 2-body currents  to $TE({\bf q},\nu)$. The transverse enhancement in both  1b and 2b currents is included in the total.
}
\vspace{-15pt}
\label{Fig4}
\end{figure*}
\vspace{15pt}

 The left panel on the bottom of  Figure  \ref{Fig3} shows a  comparison of the extracted $SL({\bf q})$ for $\carbon$ (dotted-blue curve with yellow error band) to theoretical calculations.  These include the Lovato 2016\cite{Lovato2016} "First Principle Green's Function Monte Carlo" (GFMC) calculation (solid-purple line),  Mihaila\cite{microscopic}  2000   Coupled-Clusters based calculation (AV18+UIX potential, dashed-green), and Cloet 2016\cite{Cloet} RPA calculation (RPA solid-red).
Our measurement for  $\carbon$ are in disagreement with Cloet 2016 RPA, and in reasonable agreement with  Lovato 2016 and  Mihaila 2000 except near $\bf{q}\approx$ 0.30 GeV where the contribution from nuclear excitations is significant.
 
 There is not enough  QE data  for $\oxygen$ to perform a complete analysis. 
 We find that the QE fit parameters for $\carbon$  also describe all available data on $\oxygen$.   A difference in $SL({\bf q})$  between $\carbon$ and $\oxygen$ could be the contribution of  nuclear excitations.  However as shown in Fig. \ref{Fig1}  the contributions of nuclear excitations to the $SL({\bf q})$ for $\carbon$ and $\oxygen$ are consistent with being equal.  
   
The bottom right panel  of Fig. \ref{Fig3} shows $SL({\bf q})$ for $\oxygen$  (dotted-blue  with green error band) compared to theoretical calculations.
   These include the Sobczyk  2020\cite{Coupled} "Coupled-Cluster with Singles-and Doubles (CCSD) NNLO$_{sat}$" (red-dashed line),  and   Mihaila 2000\cite{microscopic}   Coupled-Cluster  calculation with (AV18+UIX potential, dashed green line). The data are in reasonable agreement with Sobczyk  2020 and Mihaila 2000 calculations for $\oxygen$ except near $\bf{q}\approx$ 0.30 GeV where the contribution from nuclear excitations is significant.

   The $TE({\bf q},\nu)$ contribution to the QE transverse structure function $F_1({\bf q},\nu)$ for $\carbon$ is parameterized as a distorted Gaussian centered around $W\approx 0.88$ GeV and a Gaussian at $W\approx 1.2$ GeV\cite{MEC2p2h}  with $Q^2$ dependent width and amplitude. $F_1^{MEC}$=0 for $\nu<\nu_{min}$  ($\nu_{min}$=16.5 MeV). For $\nu>\nu_{min}$ it is given by:
    \begin{eqnarray}
    F_1^{MEC} &=& max((f_1^{A}+f_1^{B}),0.0)\\
    f_1^A &=& a_1Y \cdot [(W^2-W^2_{min})^{1.5} \cdot e^{-(W^2-b_1)^2/2c_1^2}] \nonumber\\
     f_1^B &=& a_2Y \cdot (Q^2+q_o^2)^{1.5} \cdot [ e^{-(W^2-b_2)^2/2c_2^2}] \nonumber\\
    Y &=& A e^{-Q^4/12.715} \frac{(Q^2+q_0^2)^2)}{(0.13380+Q^2)^{6.90679}}\nonumber\\ 
    a_1&=&0.091648,~a_2 = 0.10223\nonumber\\ 
    W^2_{min}&=& M_p^2+2M_p\nu_{min}-Q^2\nonumber
    \end{eqnarray}
   where  Q$^2$ is in units of GeV$^2$, M$_p$ is the proton mass, A is the atomic weight, $q_0^2=1.0\times10^{-4}$, $b_1$ = 0.77023, $c_1 = 0.077051+0.26795 Q^2$, $b_2$= 1.275, and $c_2$= 0.375.  %

     The parameters of the empirical model of $TE({\bf q},\nu)$ in electron scattering can be used to predict the $TE({\bf q},\nu)$  contribution in neutrino scattering\cite{Mosel}.

     A comparison between our extraction  of \rltot and \rttot and the extraction (for only three values of $\bf q$) by Jourdan\cite{Jourdan}  are shown in Fig. \ref{Fig4}. At the lowest $\bf q$ our \rltot is a somewhat lower and our \rttot is a somewhat higher (the Jourdan analysis includes data from only two experiments). 
     Also shown are  two 1-body+2-body current (1b+2b) calculations:   GFMC\cite{Lovato2016} and  "Energy Dependent-Relativistic Mean Field" (ED-RMF)\cite{ED-RMF}.
In our fit, we show each nuclear excitation with excitation energy $E_x$ at
$\nu=E_x+{\bf q}^2/2M_{C12}$ where 
$M_{C12}$ is the mass of the carbon nucleus>
In contrast, the ED-RMF calculations group all excitations in two fixed $\nu$ peaks as shown in Fig.\ref{Fig4}, and the  GFMC calculations do not show any nuclear excitations.
Both calculations 
are in reasonable agreement with our analysis in the QE region.
In both models there is enhancement of \rtqe if only 1b currents are included and additional enhancement if both 1b and 2b  currents are included.
The curves labeled 2b in Fig. \ref{Fig4} show the enhancement in \rtqe from 2b currents only [i.e. (1b+2b) minus (1b only)], while our empirical extraction of $TE({\bf q},\nu)$ shown in Fig. \ref{Fig4} models the enhancement in \rttot from all sources.

 In summary, using all available electron scattering data we extract parameterizations of the quenching of the \rlqe and the enhancement of \rtqe over a large range of $\bf q$ and $\nu$,  and obtain the best measurement of  the Coulomb Sum Rule  $SL({\bf q})$ to date. The measured $SL({\bf q})$ for $\carbon$ are inconsistent with Cloet 2016 RPA, but in reasonable agreement with  Lovato 2016 and Mihaila 2000 calculations.  The sum rule $SL({\bf q})$  for $\oxygen$ is in reasonable agreement with Sobczyk  2020 and Mihaila 2000 calculations,

  The contribution of nuclear excitations to  $SL({\bf q})$  is  significant (up to  29\%).  Theoretical studies show that at low $\bf q$ nuclear excitations are also significant in $\nu_{e,\mu}$ scattering\cite{Pandey1,Pandey2}. Therefore, nuclear excitations should be included in  both electron and $\nu_{e,\mu}$ MC generators.  Decays of excitations with $E_x$ above proton removal threshold can have a proton in the final state (which in $\nu_{e,\mu}$ experiments cannot be distinguished from QE events).

Additional comparisons of our fit to experimental data are included in the supplemental materials\cite{Supplemental}.  New precision $^{12}C$ QE data from both Hall C~\cite{HallC} (at low $\bf q$ and forward angles) and Hall A (specifically taken to examine the saturation of the CSR) are expected to be finalized soon and to further improve the separation of longitudinal and transverse cross sections via future fitting efforts. 
 This Research is supported by the U.S. Department of Energy, Office of Science, under University of Rochester grant number DE-SC0008475, and the Office of Science, Office of Nuclear Physics under contract DE-AC05-06OR23177.
    %
\section{Supplementary materials}
Additional comparisons of the universal fit to some of the measurements of  $\frac{d^2\sigma}{ d\Omega d \nu}$ on $\carbon$ and $\oxygen$.
%
\begin{figure*}[h]
\begin{center}
\includegraphics[width=7.0in,height=9.0in]{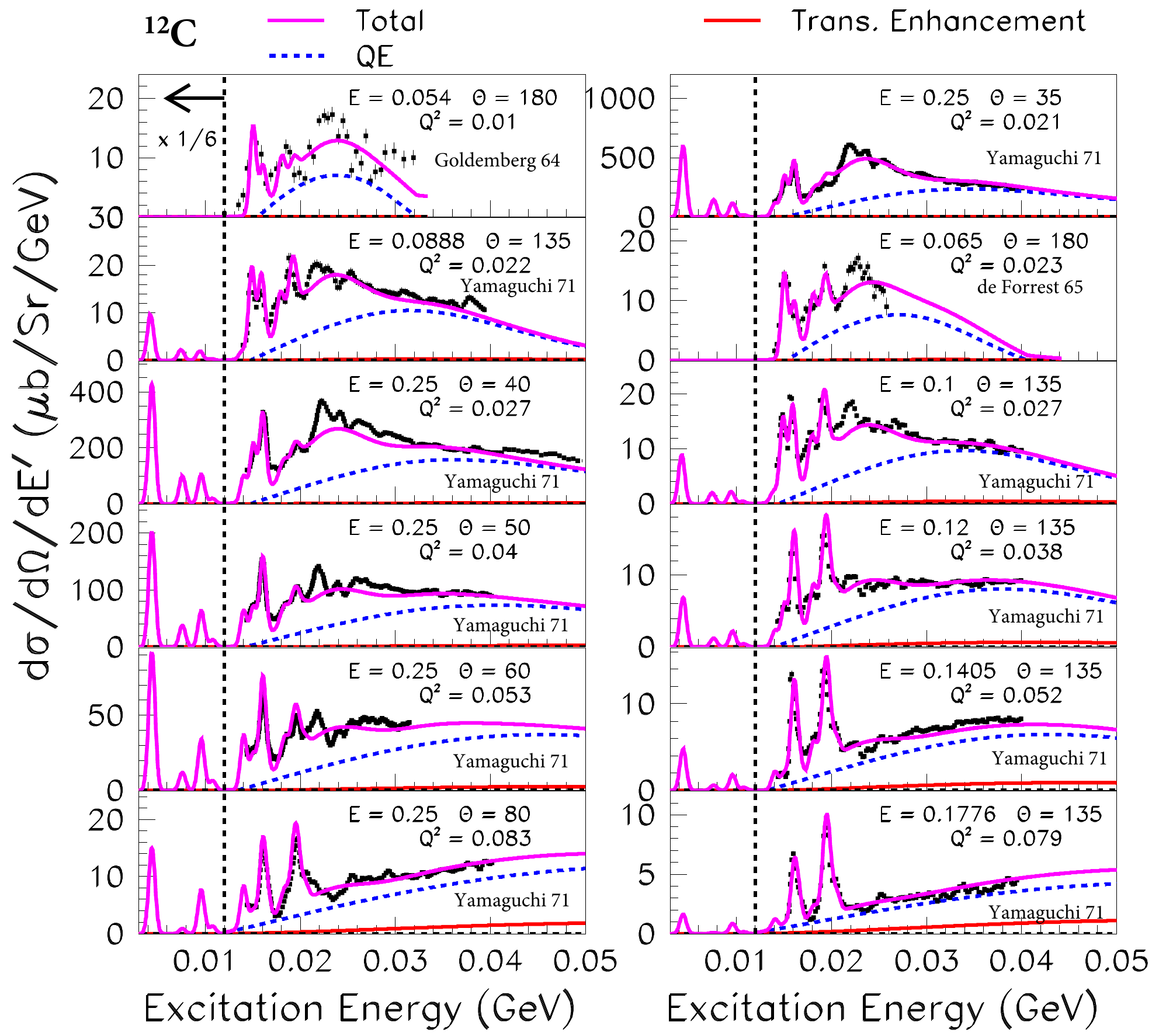}
\caption{Radiatively corrected inelastic electron scattering cross sections on $\carbon$ for excitation energies less than 50 MeV.  The  cross sections for excitation energies less than 12 MeV have been multiplied by (1/6). The pink solid line is the predicted $\frac{d^2\sigma}{d \nu d\Omega}$
from our fit which include nuclear states excitation form factor, quasilelastic scattering (dashed blue line), "Transverse Enhancement/MEC" (solid red line) and inelastic  pion production (at higher excitation energies).  Most of the data is from Yamaguchi 1971 (Phys. Rev. D3 (1971) 1750).
The data for 54 MeV and 180$^0$ are from Goldemberg 1964 (Phys. Rev. 134 (1964) B963), and the  data for 65 MeV and  180$^0$ are from deForest 1965 (Phys. Letters 16 (1965), 311).}
\label{Yamaguchi_states}
\end{center}
\end{figure*} 
%
\begin{figure*} 
\begin{center}
\includegraphics[width=6.5in, height=4.5 in]{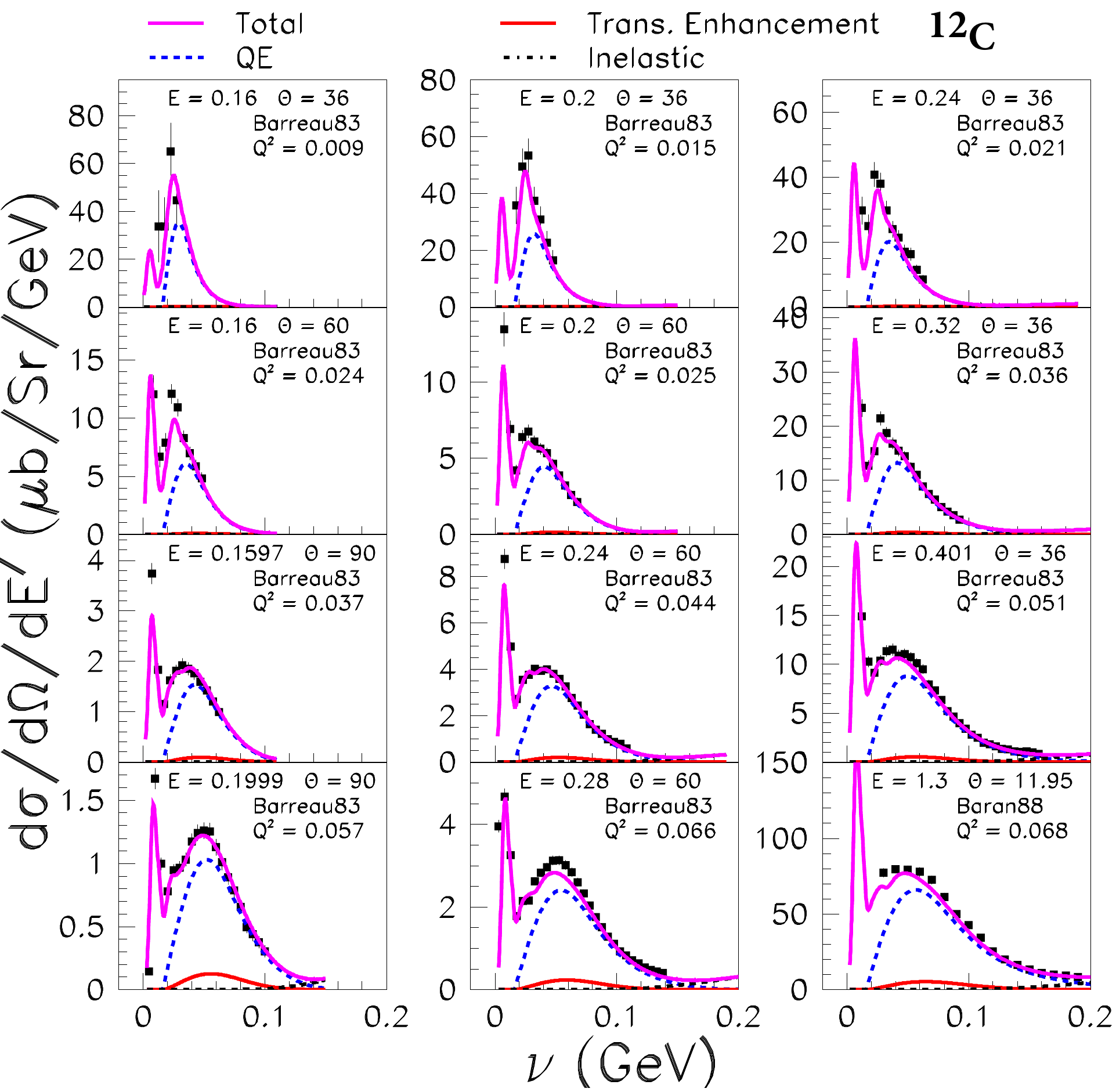}
\includegraphics[width=6.5in, height=4.5 in]{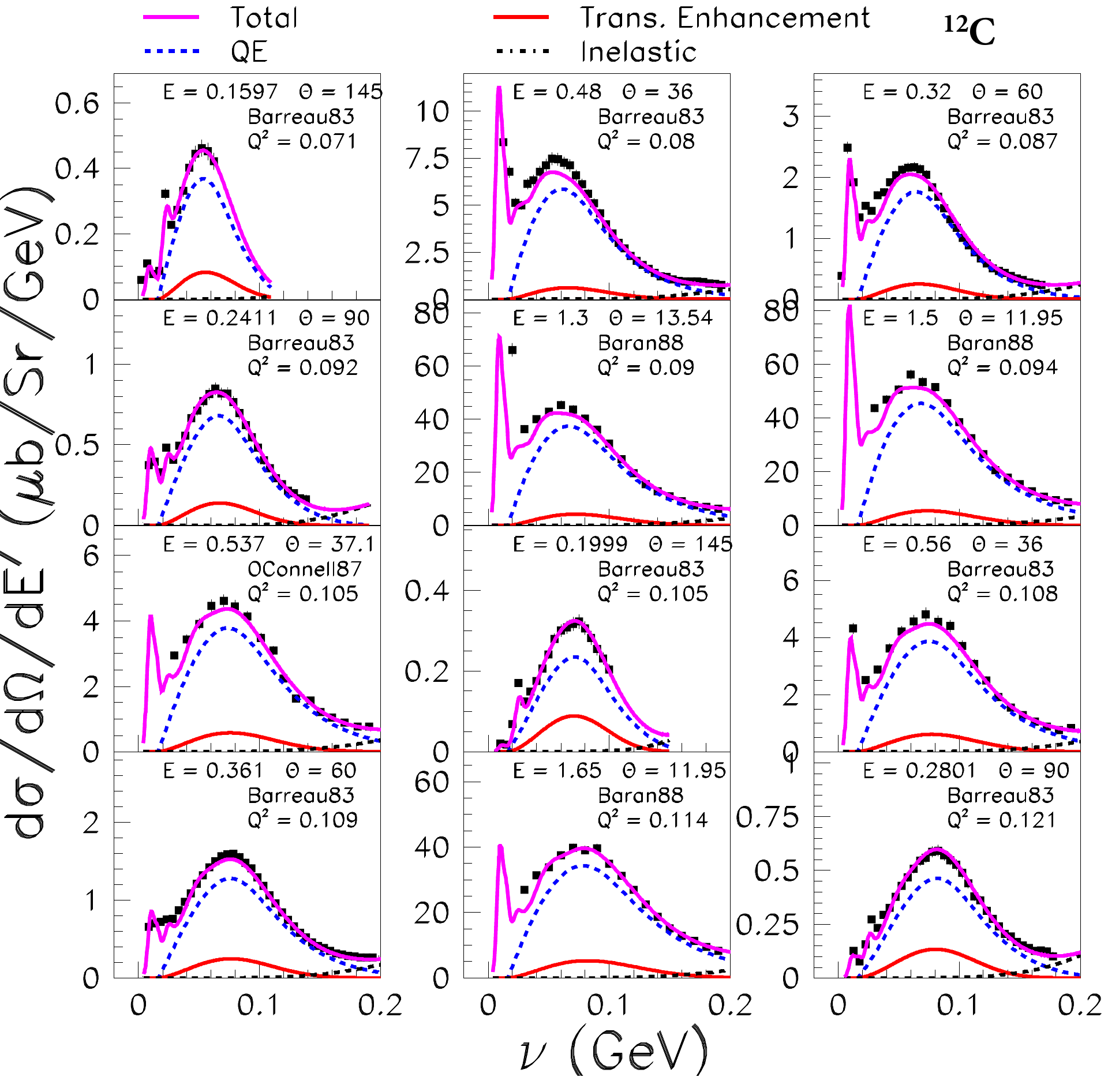}
\caption { Comparisons our fit to a subset of  electron scattering differential  cross section data with $Q^2<0.12$ GeV$^2$. 
The total  $\frac{d^2\sigma}{d \nu d\Omega}$ is shown as the solid purple line.  The dashed blue line is the QE differential cross section. The TE contribution to the QE differential cross section  is shown as the solid red line. Inelastic  pion production processes are shown as the dot-dashed black line.  The fit is in good agreement with the cross section data for both small and large angles.  The values of $Q^2$ increase from top bottom from $Q^2$=0.009 to $Q^2$=0.121 GeV$^2$. Data are from Barreau 1983 (Nucl. Phys. 402A (1983) 515) and Baran 1988 (Phys. Rev. Lett. 61 (1988) 400).
}
\label{cross-fig-1}
\end{center}
\end{figure*}

\begin{figure*}
\begin{center}
\includegraphics[width=6.5 in, height=7. in]{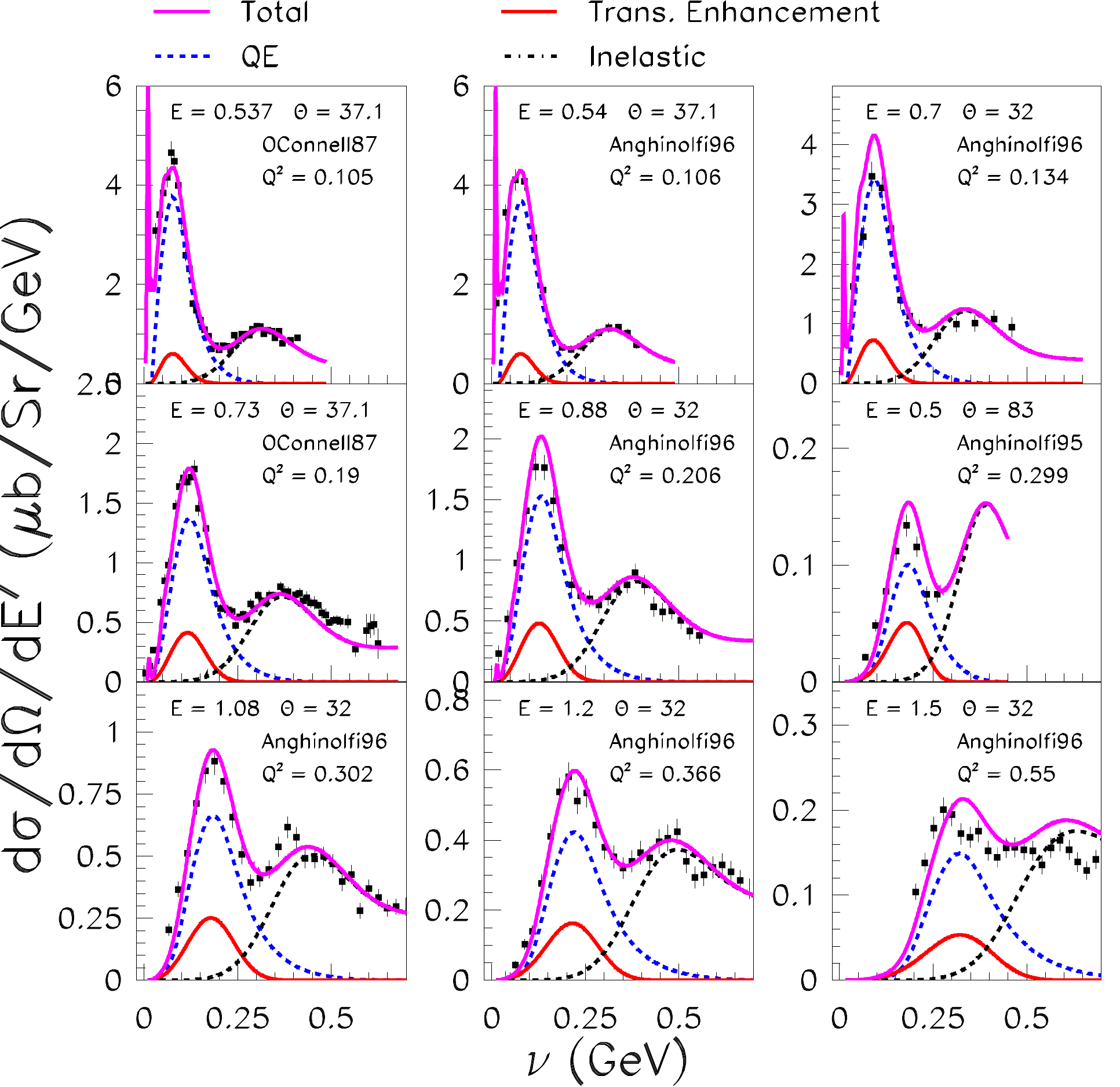}
\i\caption {Comparison of our fit (using $\carbon$ parameters) to all available $\frac{d^2\sigma}{d \nu d\Omega}$ measurements  on $\oxygen$.
Data are from O'Connell 1987 (Phys. Rev. C35 (1987) 1063), Anghinolfi 1995  
(J. Phy. G: Nucl. Part. Phys. 21 L9, 1995) and Anghinolfi 1996 (Nucl.  Phys. A602 (1996) 405.}
\label{O16_QE}
\end{center}
\end{figure*}

\begin{thebibliography}{9}
  %
\bibitem{CSR}  D. Drechsel and M M Giannini 1989 Rep. Prog. Phys. 52 1083 (eq. 7.9); T. de Forest Jr. and J.D. Walecka, Advances in Physics, 15:57, 1-109 (1966) (eq. 6.8).
\bibitem{miniboone}  A. A. Aguilar-Arevalo et al., (MiniBooNE) Phys. Rev. Lett. 100, 032301 (2008).
\bibitem{minerva}M. F. Carneiro et al., (MINERvA Collaboration),  Phys. Rev. Lett. 124, 121801 (2020).
 \bibitem{microboone1}P. Abratenko et al., (MicroBooNE),   Phys. Rev. Lett. 123, 131801 (2019).
\bibitem{microboone2} P. Abratenko et al., (MicroBooNE) Phys. Rev. Lett. 125, 201803 (2020).
   \bibitem{GEp} J.A. Caballero,  M. C. Martinez, J. L. Herraiz, J.M. Udias  Physics Letters B 688, 250 (2010).
 \bibitem{BBBA} A. Bodek, S. Avvakumov, R. Bradford, H. Budd,  Eur. Phys. J C53, 349 (2008).
\bibitem{Yamaguchi} Y. Yamaguchi et. al.,., Phys. Rev. D3, 1750 (1971).
\bibitem{Bosted}  P.E. Bosted and V. Mamyan,  arXiv:1203.2262 (2012); V. Mamyan, Ph.D. dissertation, Univ. of Virginia, 2010. 
 \bibitem{Donnelly} C. Maieron, T.W. Donnelly, I. Sick,  Phys.Rev. C65, 025502, (2002). 
 \bibitem{Amaro1}  J.E. Amaro, M.B. Barbaro, J.A. Caballero,  T.W. Donnelly, A. Molinari, and I. Sick,  Phys. Rev. C 71, 015501 (2005).
 \bibitem{Amaro2} J.E. Amaro, M.B. Barbaro, J.A. Caballero, R. Gonzalez-Jimenez, G.D. Megias, I. Ruiz Simo, J. Phys. G: Nucl. Part. Phys. 47, 124001 (2020).
 \bibitem{Megias} G.D. Megias Vazquez
 (Tesis Doctoral). Universidad de Sevilla, Sevilla (2017).
 \bibitem{Rosenfelder}  R. Rosenfelder, Ann. Phys. 128, 188 (1980).
\bibitem{Megias2} G. D.  Megias,  M. V. Ivanov, R. Gonzalez-Jimenez, M. B. Barbaro,J. A. Caballero, T. W. Donnelly,  J. M. Udias,
 Phys. Rev. D 89, 093002 (2014).
\bibitem{transverse}  A. Bodek, H. Budd, snd  M. Christy,    Eur.Phys.J.C 71, 1726 (2011).
\bibitem{archive}O. Benhar, D. Day and I. Sick, Rev. Mod. Phys. 80, 189 (2008). Quasielastic~Electron~Nucleus~Scattering~Archive.  http://discovery.phys.virginia.edu/research/groups/qes-archive/
\bibitem{resonance} Resonance-Data-Archive: https://hallcweb.jlab.org/resdata/
\bibitem{Coulomb} P. Gueye, M.  Bernheim, J. F.  Danel,  J. E. Ducret, L. Lakehal-Ayat, J. M. LeGoff,  A. Magnon et al.,
 Phys. Rev. C60, 044308 (1999).
\bibitem{FreeN}   P.E. Bosted, Phys. Rev. C 51, 409 (1995).
\bibitem{removal} Arie Bodek and Tejin Cai, Eur. Phys. J. C. (2019) 79;  Arie Bodek and Tejin Cai,  Eur. Phys. J. C 80, 655 (2020).
\bibitem{ChristyD}    P.E. Bosted and M.E. Christy, Phys. Rev. C 77, 065206 (2008);
M.E. Christy and P.E. Bosted, Phys. Rev. C 81, 055213 (2010). 
\bibitem{Barreau} P. Barreau et. al., Nucl. Phys. 402A, 515 (1983).
\bibitem{Baran} D. T. Baran et. al.,, Phys, Rev. Lett. 61, 400 1(988).
\bibitem{Jourdan} J. Jourdan, Nucl. Phys. A603, 117 (1996); J. Jourdan,
Phys. Lett. B353, 189 (1995).
\bibitem{Lovato2016} A.  Lovato, S. Gandolfi,  J. Carlson, S. C. Pieper, R. Schiavilla,
Phys. Rev. Lett. 117, 082501 (2016). (GFMC: First Principle Green's Function Monte Carlo).
\bibitem{microscopic}    Bogdan Mihaila and Jochen H. Heisenberg, Phys, Rev. Lett.  84, 1403 (2000).
\bibitem{Cloet} Ian C. Cloet, Wolfgang Bentz, Anthony W. Thomas, Phys. Rev. Lett. 116, 032701 (2016).
\bibitem{Coupled}   J. E. Sobczyk, B. Acharya, S. Bacca, and G. Hagen  Phys.Rev.C 102, 064312 (2020). 
 \bibitem{MEC2p2h} A. V. Butkevich and S. V. Luchuk, Phys.Rev.C 102,  024602 (2020).
 \bibitem{Mosel} K.  Gallmeister, U. Mosel, and J. Weil, Phys. Rev. C 94, 035502 (2016);  S. Dolan, U. Mosel, K. Gallmeister, L. Pickering, and S. Bolognesi Phys. Rev. C 98, 045502 (2018).
%
\bibitem{ED-RMF}  T. Franco-Munoz, R. Gonzalez-Jimenez, and J.M. Udias, arXiv:2203.09996v1 (2022).(ED-RMF: Energy Dependent-Relativistic Mean Field).
\bibitem{Pandey1} V. Pandey, N. Jachowicz, T. Van Cuyck, J. Ryckebusch,  and M. Martini, Phys. Rev. C 92, 024606 (2015),
\bibitem{Pandey2} M. Martini, N. Jachowicz, M. Ericson, V. Pandey, T. Van Cuyck, and N. Van Dessel, Phys. Rev. C 94, 015501 (2016); 
V.  Pandey,   N. Jachowicz, M. Martini, R. Gonzalez-Jimenez, J. Ryckebusch, T. Van Cuyck, and N. VanDessel, 
 Phys. Rev. C 94, 054609 (2016). 
  \bibitem{HallC} 
  Jlab Hall C E04-001 Spokespersons: A. Bodek, M. E. Christy, C. Keppel;  A. Bodek arXiv:hep-ex/0411044; QE data at $\bf q$ values of 0.225, 0.267, 0.337, 0.395, 0.447, 0.571, 0.854 GeV.
   \bibitem{HallA} 
  Jlab Hall A E05-110 Spokespersons : Jian-Ping Chen, Seonho Choi, Zein-Eddine Meziani. QE data at 0.55 GeV $<{\bf q}<$ 1.0 GeV.
   https://www.jlab.org/exp prog/proposals/05/PR05-110.pdf
   
   %
 \bibitem{Supplemental}
Supplemental materials [URL will be inserted by publisher] 
\end{thebibliography}
\end{document}